\documentclass[journal]{IEEEtran}
\usepackage{ifpdf} 
\usepackage{cite} 
\usepackage{amsmath} 
\usepackage{color} 
\usepackage{array} 
\usepackage{url} 
\usepackage{multirow}
\usepackage{array}
\usepackage{xcolor}
\usepackage{amssymb}
\usepackage{tikz}
\usepackage{caption}
\usepackage{physics}
\usepackage{graphicx}
\usepackage{yhmath}
\usepackage{mathdots}
\usepackage{MnSymbol}
\usepackage{lipsum}
\usepackage{svg}
\usepackage{float}
\usepackage{bbm}
\usepackage[export]{adjustbox}
\usepackage{amssymb}
\usepackage{hyperref}
\usepackage[english]{babel}

\allowdisplaybreaks

\usepackage[caption=false,font=normalsize,labelfont=sf,textfont=sf]{subfig}

\usepackage{eqparbox}
\usepackage{arydshln}
\usepackage{float}

\newtheorem{proposition}{Proposition}

\begin{document}
\title{APECS: Adaptive Personalized Control System Architecture}

\author{Marius~F.~R.~Juston$^{1}$,
        Alex~Gisi$^{2}$,
        William~R.~Norris$^{3}$,
        Dustin~Nottage$^{4}$,
        Ahmet~Soylemezoglu$^{4}$

\thanks{Marius F. R. Juston$^{1}$ is with The Grainger College of Engineering, Industrial and Enterprise Systems Engineering Department, University of Illinois Urbana-Champaign, Urbana, IL 61801-3080 USA (e-mail: mjuston2@illinois.edu).}

\thanks{Alex Gisi$^{2}$ is with The Grainger College of Engineering, Electrical and Computer Engineering Department, University of Illinois Urbana-Champaign, Urbana, IL 61801-3080 USA (e-mail: wrnorris@illinois.edu).}

\thanks{William R Norris$^{3}$ is with The Grainger College of Engineering, Industrial and Enterprise Systems Engineering Department, University of Illinois Urbana-Champaign, Urbana, IL 61801-3080 USA (e-mail: wrnorris@illinois.edu).}

\thanks{Construction Engineering Research
Laboratory$^{4}$, U.S. Army Corps of Engineers
Engineering Research and Development
Center, IL, 61822, USA}

\thanks{This research was supported by the U.S. Army Corps of Engineers Engineering Research and Development Center, Construction Engineering Research Laboratory.}
}

\markboth{IEEE TRANSACTIONS ON XXX XXXX, XXX, XXX, September~2023}%
{Shell \MakeLowercase{\textit{et al.}}: Bare Demo of IEEEtran.cls for IEEE Journals}


\maketitle

\begin{abstract}
This paper presents the Adaptive Personalized Control System (APECS) architecture, a novel framework for human-in-the-loop control. An architecture is developed which defines appropriate constraints for the system objectives. A method for enacting Lipschitz and sector bounds on the resulting controller is derived to ensure desirable control properties. An analysis of worst-case loss functions and the optimal loss function weighting is made to implement an effective training scheme. Finally, simulations are carried out to demonstrate the effectiveness of the proposed architecture. This architecture resulted in a 4.5\% performance increase compared to the human operator and 9\% to an unconstrained feedforward neural network trained in the same way.
\end{abstract}


\IEEEpeerreviewmaketitle

\section{Introduction} \label{s_Intro}

\IEEEPARstart{H}{uman-in-the-loop} control allows robots to address problems inaccessible to full autonomy for reasons of liability, trust, or safety. Examples include advanced driver assistance systems \cite{saleh2013shared, wang2020decision, nguyen2016driver}, tele-operated surgery devices with haptic feedback \cite{Enayati2018, Xiong2017, Su2022}, assistive robots for those with impairments \cite{nguyen2013shared, song2024driving, wang2014adaptive}, and aerial/underwater vehicle control systems \cite{franchi_shared_2012, lee2013semiautonomous, brantner2021controlling}. In such applications, humans use high-level reasoning skills to provide intention and the ability to adapt to unforeseen or unfavorable operating circumstances.

Since the human is directly inserted in the control loop, his skill in operating the machine will have a significant impact on the outcome of the task. Expert operators are expensive and may not always be available. Training novice operator to become experts can be a lengthy process. Data collected from novices learning a compensatory tracking task as part of aircraft pilot training indicates initial control skill acquisition can be very slow, continuing over many dozens of training runs \cite{mulder_manual_2018}. 

Instead of changing the operator to better use the system, the system can be changed to be more appropriate for the operator. Automatically adapting the parameters of a system to improve some measure of performance is known as human-in-the-loop optimization. Human-in-the-loop optimization has been applied to medical assistance devices like exoskeletons \cite{Haufe_Wolf_Riener_2020, Pang_Li_Ding_Tang_Luo_Xiang_2023, Slade_Kochenderfer_Delp_Collins_2022, Xu_Liu_Chen_Yu_Yan_Yang_Zhou_Yang_2023} and prostheses \cite{Garcia_Rosas_Tan_Oetomo_Manzie_Choong_2021, Wen_Si_Brandt_Gao_Huang_2020}. Algorithms used include evolutionary algorithms, surrogate optimization, gradient-based methods, and reinforcement learning, with the problems ranging from 2-22 parameters \cite{slade_human_loop_2024}. The optimization criterion is typically selected to measure the device's ease of use, such as minimizing metabolic expenditure \cite{Haufe_Wolf_Riener_2020, Slade_Kochenderfer_Delp_Collins_2022} or muscle fatigue \cite{Pang_Li_Ding_Tang_Luo_Xiang_2023}. The result is devices which offer personalized assistance. A high-level overview of human-in-the-loop optimization is \cite{slade_human_loop_2024}. An in-depth survey of human-in-the-loop optimization for devices (exoskeletons/exosuits/prostheses) which improve locomotion is \cite{zhang2024closing}. 



To the authors' knowledge, the earliest example of human-in-the-loop optimization was the dissertation of Norris \cite{william_norris_design_2001}, which developed a method for online adaptation of the steering gain in a skid-steer vehicle. In that work and \cite{william_r_norris_virtual_2001, william_r_norris_novel_2002, w_r_norris_design_2003}, the ``virtual design tools" were developed, a modular toolset for systematically modeling and incorporating human behavior directly into the design process. The toolset includes the \textit{virtual machine}, a dynamic system model that incorporates adaptable design parameters; the \textit{virtual designer}, which uses an error signal to optimize the adaptable parameters; and the \textit{virtual operator}, a control system with human-like qualities (e.g. fuzzy controller) which serves as a consistent operator during the design process so improvement can be demonstrated ceteris paribus. When the optimization is applied to the system parameters, the \textit{virtual modulation surface} (VMS) is obtained, which describes how the human inputs are mapped to the fixed system controls. The work \cite{william_norris_design_2001} implemented an offline technique to implement the virtual design tools in one dimension, resulting in the \textit{virtual modulation curve} (VMC). The framework was applied to a wheel loader, and resulted in improved trajectory tracking on a Society of Automotive Engineers steering test course. However, the system was a proof of concept, and incorporating higher dimensional input/output or time-varying gains was not addressed.

In the present work, a similar structure is used to implement a general VMS (a multiple-input, multiple-output mapping). Accordingly, we address the offline open-loop optimization of an arbitrary control system for improved operator performance by proposing the Adaptive Personalized Control System (APECS) architecture, training procedure, and application. 


Informally, the objective of APECS is to modulate operator input in order to improve performance in tracking tasks without a priori knowledge of the plant. To maintain operator understanding of the system, the modulation should not be unexpected, for example, it should not change the control sign or exhibit large variations from a small input. 

Various methods for designing Lipschitz-constrained networks have been explored in the literature, but typically from the perspective of robustness certification against adversarial attacks, especially in the context of image classification \cite{Tsuzuku2018, Huang2021, leino2021globallyrobust, Fazlyab2019, Trockman2021, Prach2022, Miyato2018, Meunier2022, Latorre2020, Gouk2018, Bartlett2017}. By bounding the slope of the loss landscape around a datapoint, one can be certain there is no adversarial example that changes the classification via a small input perturbation \cite{Tsuzuku2018}. On the other hand, this is also a highly desirable property for a vehicle controller because operators intuitively expect small changes in their input to result in small changes to the system output. The global Lipschitz constant can also be used for stability analysis of the closed-loop system \cite{Fazlyab2019}.

Recently, \cite{Araujo2023} offered a common theoretical framework for generating 1-Lipschitz layers for standard residual and feed-forward neural networks. The authors view the neural network as a Lur'e system and apply an LMI constraint to obtain the 1-Lipschitz property. If the neural network is not defined using inherently L-Lipschitz layers, an alternative is to approximate the Lipschitz constant using computationally expensive approximation \cite{Tsuzuku2018, leino2021globallyrobust, Huang2021}. We follow the approach of \cite{Araujo2023} to implement an L-Lipschitz neural network, which forms the basis of the controller. However, we apply nonlinear transformations to the network output to achieve the desired controller properties. Hence, we must take additional steps to derive bounds for the Lipschitz constant of the complete control system. 

The main contributions of the work are
\begin{itemize}
    \item Architecture design for offline optimization of human input for arbitrary plant control
    \item Novel approach for elementwise sector bounding neural network output
    \item Derivation of an optimal weighting for the loss function to ensure balancing between optimal and human operators
\end{itemize}

The paper is organized as follows. In Section \ref{sec:notation}, the mathematical notation is described. In section \ref{sec:problem-formulation}, the system's requirements are stated formally. In section \ref{sec:architecture}, a system is proposed which satisfies the requirements.

\section{Preliminaries}
\label{sec:notation}
A function $f: \mathbb{R}^m \to \mathcal{D}$, where $\mathcal{D} \subseteq \mathbb{R}^n$, is globally L-Lipschitz if and only if $\lVert f(x) - f(y) \rVert_2 \leq L \lVert x - y \rVert_2$ for all $x, y \in \mathcal{D}$. The Lipschitz constant of $f(\cdot)$ is the smallest such $L$. Given a vector $x$, the matrix $\text{diag}(x)$ is that with the elements of $x$ on the diagonal and zeros elsewhere. The vector operator $\otimes$ denotes the elementwise multiplication $A \otimes B = C$, where $C_{ij} = A_{ij} B_{ij}$.

\section{Problem Formulation} \label{sec:problem-formulation}
A human operates an electromechanical system $\dot{y} = f(y, u)$ to track a reference signal $\bar{y}$. The system control input is assumed
\begin{equation}
    u = \hat{x}(x, e, \mathcal{E}),
    \label{eq:u-all}
\end{equation}
where $x \in D = [-1, 1]^{n_x}$ is the human input, $e = y - \bar{y}$, and $\mathcal{E} \in \mathbb{R}^{n_\mathcal{E}}$ are environment parameters observed through sensors.


The problem is to determine the mapping $\hat{x}(x, e, \mathcal{E})$ between the human command and the system input which at each time satisfies:
\begin{enumerate}
    \renewcommand{\labelenumi}{R\arabic{enumi})}
    \item $\hat{x} \in [-1, 1]^{n_x}$
    \item $x = 0 \iff \hat{x} = 0$
    \item $\text{sign}(x) = \text{sign}(\hat{x})$
    \item $\lVert \hat{x}(x_1, \cdot, \cdot) - \hat{x}(x_2, \cdot, \cdot) \rVert_2 \leq L \lVert x_1 - x_2 \rVert_2, \ \forall x_1, x_2 \in \mathcal{D}$
    \item optimal with respect to a task running cost $\mathcal{L} = J(x, \hat{x}, \bar{x})$
\end{enumerate}

\section{APECS Architecture}
\label{sec:architecture}

Consider the control signal
\begin{align}
    \hat{x}(x, \mathcal{E}, e) &= s \left( p ( g_\theta(x, \mathcal{E}, e) ) \otimes x \right),
    \label{eqn:APECSMain}
\end{align}
where $s(\cdot): \mathbb{R}^{n_x} \to [-1, 1]^{n_x}$ is an elementwise sector-bounded saturation function satisfying $s(0) = 0$, $p(\cdot): \mathbb{R}^{n_x} \to [0, \infty)^{n_x}$ is an element-wise positive definite function, and $g_\theta(x, \mathcal{E}, e): [-1, 1]^{n_x} \times \mathbb{R}^{n_\mathcal{E}} \times \mathbb{R}^{n_e} \to \mathbb{R}^{n_x}$ is a feed-forward neural network with parameter vector $\theta$.

By construction of $s$, the proposed $\hat{x}$ trivially satisfies R1-R3. Section \ref{sec:lipschitzConstraint} will derive constraints which satisfy the Lipschitz condition R4. Section \ref{sec:training} will demonstrate a training method for $g_\theta$ to accomplish R5.

\begin{figure}[t]
    \centering
    \includegraphics[width=1\linewidth]{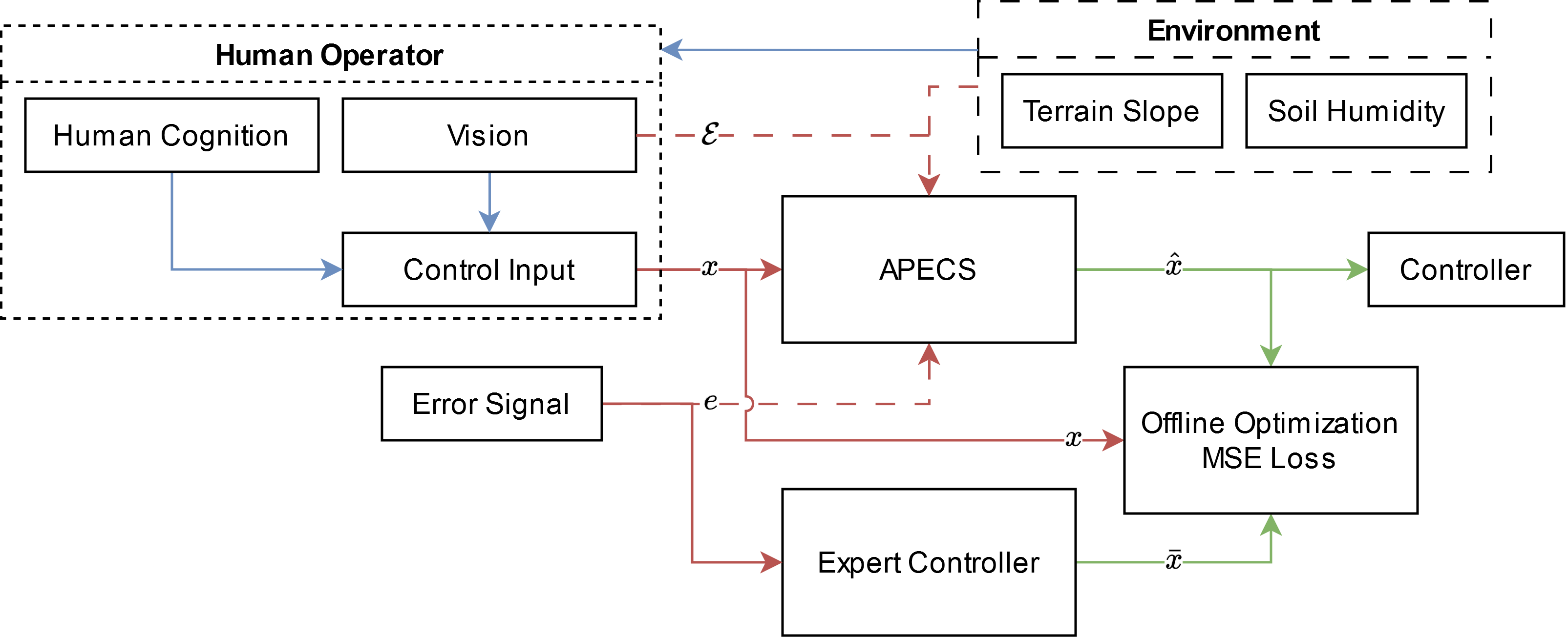}
    \caption{APECS Architecture}
    \label{fig:APECSArchitecture}
\end{figure}

Figure \ref{fig:APECSArchitecture} shows how the system will be trained to minimize $J(x, \bar{x}, \hat{x})$, where the Offline Optimization block updates $\theta$ according to an optimization scheme.

\begin{figure}[b]
    \centering
    \includegraphics[width=1\linewidth]{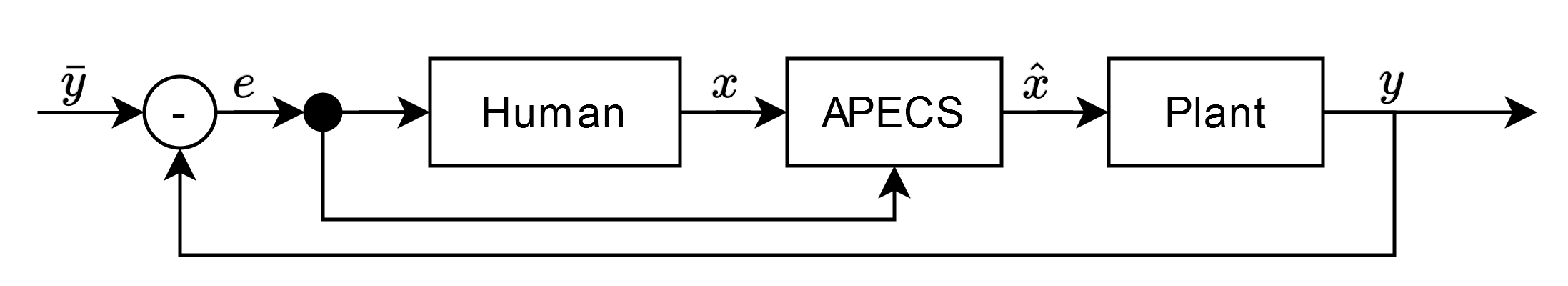}
    \caption{APECS Process Control Diagram}
    \label{fig:APECSProcessDiagram}
\end{figure}

Figure \ref{fig:APECSProcessDiagram} shows the feedback control structure. Notice no knowledge of the plant is needed to train or deploy the control architecture.

\section{Lipschitz Constraint}
\label{sec:lipschitzConstraint}

A desired property for the control signal is to enforce a Lipschitz bound with respect to the human input. The Lipschitz property on a network can help with its robustness to noise \cite{Tsuzuku2018} and provides uniqueness conditions on the generated controller output \cite{Derrick1976}.

The implementation of a neural network to be L-Lipschitz has been accomplished with various methods \cite{Tsuzuku2018, Huang2021, leino2021globallyrobust, Fazlyab2019, Trockman2021, Prach2022, Miyato2018, Meunier2022, Latorre2020, Gouk2018, Bartlett2017}. In  \cite{Araujo2023}, the authors gave a network architecture that unifies the approaches for generating Lipchitz neural networks.

Accordingly, we assume the neural network $g_\theta(\cdot)$ has been implemented as L-Lipschitz. However, due to the nonlinearity of $\hat{x}(\cdot) = s(\cdot)$, the Lipschitz constant of the resulting controller does not proportionally reflect the Lipschitz constant of the neural network. We now derive bounds on the relationship between them. To do so we must specify $s(\cdot), p(\cdot)$ which satisfy R1-R3. Let
\begin{align}
    s(x) &= \max(-1, \min(1, x)) \hspace{0.2cm} \text{(clip)}, \\
    p(x) &= \ln(1 + e^x)) \hspace{0.2cm} \text{(softplus)}.
\end{align}
The softplus function is described in \cite{Dugas2000}. For the positive definitive function $p(\cdot)$ we also enforce the constraint that it is strictly monotonically increasing and it's first derivative is also monotically increasing. As such if $x < y$ such that $p(x) < p(y)$ and $0 \leq \frac{d}{dx}p(x) < \frac{d}{dy}p(y)$ then this will be a valid positive function $p(\cdot)$.

\subsection{Lipschitz bounding}

The neural network satisfies 
\begin{equation*}
    g_\theta(z) \leq \hat{g}_\theta(z) := L_\theta \abs{x} + b_\theta    
\end{equation*}
where $L_\theta$ is the (known) Lipschitz constant of the neural network and $b_\theta=g_\theta(0)$ is the bias. For convenience, let $z = \{ x, \mathcal{E}, e \}$. The Lipschitz constant with respect to the human input is the maximum value of the partial derivative, with $\abs{x} \leq c$ and the maximum gradient $L_\theta$ satisfies $-L_\theta \leq \norm{g'(z)} \leq L_\theta$.
\begin{align*}
    \hat{x}(z) &= s(p(g_\theta(z)) \otimes x) \\
    \frac{\partial \hat{x}(z) }{\partial x} =& \begin{cases}
    p'(g_\theta(z)) g'_{\theta}(z) \otimes x +  p(g_\theta(z)) & \abs{p(g_\theta(z)) \otimes x} < 1 \\
    0 & \text{Otherwise}
    \end{cases} \\
    \leq&  p'(L_\theta \abs{x} + b_\theta) L_\theta \otimes \abs{x} +  p(L_\theta \abs{x} + b_\theta) \\
    \leq& p'(L_\theta c + b_\theta)  L_\theta c +  p(L_\theta c  + b_\theta)
\end{align*}
As such the Lipschitz constant of the transformed network is thus, $L_p = p'(L_\theta c + b_\theta)  L_\theta c +  p(L_\theta c  + b_\theta)$. The transformed network can thus be scaled in the following way, 
\begin{align*}
    y(z) &= s\left(\frac{1}{L_p} p(g_\theta(z)) x \right).
\end{align*}
resulting in a 1-Lipschitz system. 
To transform the function \ref{eqn:APECSMain} into an $L_t$-Lipschitz function, a trainable scaling factor, $L_t > 0$, can be utilized to train the network. To ensure the positive definiteness of the scaling factor, we can reparametrized the parameter $L_t$ as $L_t = e^\alpha$:
\begin{align*}
     y(z) &= s\left(\frac{e^{\alpha}}{L_p} p(g_\theta(z)) x \right).
\end{align*}


\section{Training} 
\label{sec:training}

In this section, we present a training scheme for the network derived in section V. The objective is to ensure that the resulting system respects the inputs from the human operator, $y_h$, to reflect the personalized style while also integrating the expert operator's reference signals $y_e$. To facilitate this, a variable loss function is constructed. 


Recall $\text{dim}(x) = n_x$. Let the the dataset be a collection of $N$ points $\mathcal{D} = \{z_i \ | \ i \in \{1, \cdots , N\}\}$. It is assumed that as $N \to \infty$, the data points are sampled uniformly and uniquely along the whole dataset dimension.

Let the losses corresponding to control signal difference from the human and expert operators be 
\begin{align*}
    \mathcal{L}_h &= \frac{1}{N n_x}\sum_{\{z\} \in \mathcal{D}}\norm{\hat{x}(z) - x}_2^2, \\
    \mathcal{L}_e &= \frac{1}{N n_x} \sum_{\{z\} \in \mathcal{D}}\norm{\hat{x}(z) - \bar{x}}_2^2,
\end{align*}
respectively. This is the generic MSE loss function implemented in e.g. PyTorch \cite{MSEPyTorch}. Then, the combined loss function
\begin{equation*}
    \mathcal{L} = \gamma \mathcal{L}_h + (1-\gamma) \mathcal{L}_e,
    \label{eq:loss}
\end{equation*}
allows us to dynamically tune $\gamma \in [0, 1]$ to determine the relative importance of mimicking the human or expert operator. 

Given that its inputs and outputs are constrained to be $ [-1, 1]$, we can define the system's maximum error.

\subsection{Maximum expert operator loss}

We first examine the maximum error function of the expert operator $\mathcal{L}_e$. We assume the worst-case scenario where 
\begin{equation*}
    \bar{x} = -\alpha \text{sign}(x), \hspace{0.2cm} \alpha \in [0, 1].
\end{equation*}
The parameter $\alpha$ can be set as an appropriate lower bound for the output of the expert controller. By the construction of $\hat{x}$ from sections \ref{sec:architecture} and \ref{sec:lipschitzConstraint}, we assume $y(z)$ is $L_t$-Lipchitz, centered at $0$, and is thus bounded by $\abs{y(z)} \leq L_t x$. 

Note the case where $L_t \leq 1$ is distinct from $L_t > 1$, because in the latter case, we can take $y(x)=1$ for $x \geq 1/L_t$ due to the output of $y(\cdot)$ being restricted to $[-1, 1]$. For the unsaturated case where $L_t \leq 1$:
\begin{align*}
    \mathcal{L}_e &= \frac{1}{N n} \sum_{\{z\} \in \mathcal{D}}\norm{y(z) - y_e}_2^2 \\
    &\leq \frac{1}{N n} \sum_{\{z\} \in \mathcal{D}}\norm{y(z) - (- \alpha\text{sign}(x))}_2^2 \\
    &\leq  \frac{1}{N n}\sum_{\{x, \mathcal{E}, e \} \in \mathcal{D}} \norm{L_t x  + \alpha\text{sign}(x)}^2_2.
\end{align*}
Taking limits,
\begin{align*}
    \lim_{N \to \infty} \mathcal{L}_e &= \lim_{N \to \infty}  \frac{1}{N n}\sum_{\{x, \mathcal{E}, e \} \in \mathcal{D}} \norm{L_t x + \alpha\text{sign}(x)}_2^2 \\
    &=  \int_{0}^{1} \left(L_t^2 x^2 + 2 L_t \alpha x + \alpha^2\right) dx \\
    &= \alpha^2 + \alpha L_t + \frac{L_t^2}{3}.
\end{align*}


For the saturated case where $L_t > 1$, we use the fact that the function $y(z)$ is clamped between $[0, 1]$. Thus for $\frac{1}{L_t} \leq x \leq 1$, $\abs{y(z)} = 1$. Then
\begin{align*}
    \mathcal{L}_e &= \frac{1}{N n} \sum_{\{z\} \in \mathcal{D}}\norm{\max(-1, \min(1, y(z))) - y_e}_2^2
\end{align*}
It follows
\begin{align*}
    \lim_{N \to \infty} \mathcal{L}_e
    &= \lim_{N \to \infty}  \frac{1}{N n}\sum_{\{x, \mathcal{E}, e \} \in \mathcal{D}} \|\max(-1, \min(1,L_t x)) \nonumber \\
    &\quad\quad\quad\quad\quad\quad\quad\quad\quad\quad\quad\quad\quad\quad\quad + \alpha\text{sign}(x)\|_2^2 \\
    &=   \int_{0}^{\frac{1}{L_t}} \left(L_t^2 x^2 + 2 L_t \alpha x + \alpha^2 \right) dx  +  \int_{ \frac{1}{L_t}}^{1} (1 + \alpha)^2 dx\\
    &= \frac{3 \alpha  (\alpha +1)+1}{3 L_t} +  (\alpha +1)^2 \left(1-\frac{1}{L_t}\right) \\
    &= (\alpha +1)^2-\frac{\alpha +\frac{2}{3}}{L_t}.
\end{align*}

As such, the worst-case expert loss function becomes:
\begin{align}
     \hat{\mathcal{L}}_e &= \begin{cases}
          \alpha^2 + \alpha L + \frac{L^2}{3}, &  0 \le L \le 1 \\
          (\alpha +1)^2-\frac{\alpha +\frac{2}{3}}{L_t}, & L \ge 1 
     \end{cases}.
     \label{eq:worst-loss-expert}
\end{align}
This loss error will be loose; a tighter bound could be achieved by assuming a Lipschitz constraint on the expert operator. However, this increases the number of cases that would need to be computed and, as such, was not derived. Future work could derive the tighter bound.

\subsection{Maximum human operator loss}


The maximum human error for the system would be if the output of the neural network $g(z)$ is a constant function returning $-\infty$ or $\infty$, returning $y(z) = 1$ and $y(z) = 0$ respectively. Either of these scenarios would function as the worst-case output since the outputs are symmetric around $y(z) = x$.

Case $L_t \leq 1$:
\begin{align*}
     \mathcal{L}_h &=  \frac{1}{N n}\sum_{\{z\} \in \mathcal{D}}\norm{y(z) - y_h}_2^2 \\
     &\leq  \frac{1}{N n}\sum_{\{x, \mathcal{E}, e \} \in \mathcal{D}}\norm{L_t x  - x}_2^2.
\end{align*}
Taking limits,
\begin{align*}
      \lim_{N \to \infty} \mathcal{L}_h &= \lim_{N \to \infty} \frac{1}{N n}\sum_{\{z\} \in \mathcal{D}}\norm{(L_t - 1) x}_2^2 \\
     &= \int_{0}^{1}(L_t - 1)^2 \norm{x}_2^2dx \\
     &= \frac{(L_t - 1)^2 }{3}.
\end{align*}

Case $L_t > 1$:
\begin{align*}
     \mathcal{L}_h &=  \frac{1}{N n}\sum_{\{z\} \in \mathcal{D}}\norm{\max(-1, \min(1,y(z))) - y_h}_2^2 \\
     &\leq  \frac{1}{N n}\sum_{\{x, \mathcal{E}, e \} \in \mathcal{D}}\norm{\max(-1, \min(1,L_t x))  - x}_2^2.
    \end{align*}
Taking the limit of,
\begin{align*}
    \lim_{N \to \infty} \mathcal{L}_h
     &= \lim_{N \to \infty} \frac{1}{N n}\sum_{\{x, \mathcal{E}, e \} \in \mathcal{D}}\norm{\max(-1, \min(1,L_t x))  - x}_2^2 \\
     &= \int_{0}^{\frac{1}{L_t}}(L_t - 1)^2 x^2 dx + \int_{\frac{1}{L_t}}^{1}(1 - x)^2 dx\\
     &= \frac{(L_t - 1)^2}{3 L_t^2}.
\end{align*}
Where, as expected the error when $\lim_{L_t \to 0}\frac{(L_t - 1)^2 }{3} = \lim_{L_t \to \infty}\frac{(L_t - 1)^2}{3 L_t^2} = \frac{1}{3}$.

As such, the constrained human loss function becomes:
\begin{equation}
         \hat{\mathcal{L}}_h =  \begin{cases}
           \frac{(L_t - 1)^2}{3 L_t^2} &  0 \le L \le 1 \\
          \frac{(L_t - 1)^2 }{3}, & L \ge 1 
     \end{cases}. \label{eq:worst-loss-human}
\end{equation}
\subsection{Equal weighting}

We can use the bounds derived from previous sections to weigh the linear combination of the losses. The weight factor $\gamma$ ensures the system does not tend too heavily to a single objective and tries to maintain characteristics that fit both operators. Accordingly, we compute the $\gamma$ in eq. \ref{eq:loss} which provides equal loss weights. From equations \ref{eq:worst-loss-expert} and \ref{eq:worst-loss-human}, we see the loss magnitudes obtain a maximum as $\lim_{L_t \to \infty}$, which become $(\alpha + 1)^2$ and $\frac{1}{3}$ respectively. To obtain equal weight scaling, we solve
\begin{equation*}
    \gamma^* \frac{1}{3} = (1 - \gamma^*) (\alpha + 1)^2,
\end{equation*}
obtaining
\begin{equation}
    \gamma^* = \frac{3 (\alpha +1)^2}{3 \alpha  (\alpha+2)+4}.
    \label{eqn:optimal_gamma}
\end{equation}
%
%

An initial guess of the optimal Lipschitz constant can be estimated by minimizing the total loss function $\mathcal{L}$ with respect to the $L_t$-Lipchitz constant.
Accordingly, we compute
\begin{align*}
    \frac{\partial \mathcal{L}}{\partial L_t} &= \begin{cases}
 -\alpha \gamma+\alpha -\frac{2 \gamma }{3}+\frac{2 L_t}{3}, & 0<L_t<1 \\
 -\frac{1}{3} (3 \alpha +2) (\gamma -1), & L_t=1 \\
 \frac{(2-3 \alpha  (\gamma -1)) L_t-2 \gamma }{3 L_t^3}, & L_t>1 
\end{cases},
\end{align*}
from which we can see that the minimum can be computed as $L_t = \frac{3}{2} \alpha (\gamma  - 1) + \gamma$ as long as the following condition holds:
\begin{align}
\left(\alpha <\frac{2 \gamma }{3-3
   \gamma }\land 0 <  \gamma \leq \frac{3}{5}\right)\lor \left(\gamma > \frac{3}{5}\right). \label{eqn:gammaConditions}
\end{align}
If the conditions above are unmet, the $L_t$ should be set close to zero for the initial $L_t$. This is because the computed $L_t$ would otherwise result in a negative number. It should be mentioned that because the case for $\mathcal{L}_e$ assumes the worst case, where the optimal controller performs the exact opposite of the expected output, these initializations are very loose. It is recommended never to initialize $L_t = 0$ as this would cause gradient and training issues. Instead, an initial value of $L_t = \frac{1}{2}$ is recommended when starting the system.
\begin{proposition}
    The $\gamma^*$, defined as \ref{eqn:optimal_gamma}, satisfies the conditions in \ref{eqn:gammaConditions} for all $\alpha$. 
\end{proposition}
\par
The proof is in Appendix \ref{sec:GammaConditionsOp}.

\section{Experiment}

\begin{figure}[!ht]
    \centering
    \includegraphics[width=1\linewidth]{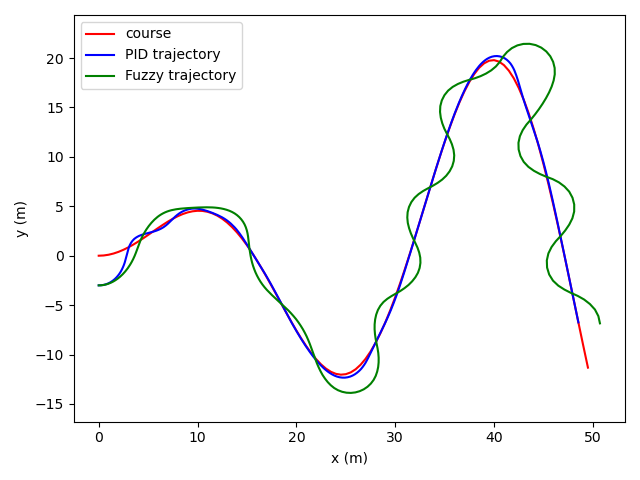}
    \caption{Default Controller trajectory following}
    \label{fig:InitialController}
\end{figure}

Simulation results were used to verify and experiment with this novel model formulation. The example human operator was a trained Fuzzy Inference system meant to act as the imperfect human operator \cite{Juston2023}, and the expert controller used Pure Pursuit Steering and PID speed control \cite{Sakai2018}. The task is trajectory following. Figure \ref{fig:InitialController} demonstrates the initial output of the expert and human operators. The human operator is intentionally tuned for poor tracking performance, to represent a novice operator.

We train three controllers. The first is a standard feed-forward neural network. The second is the APECS with the Lipschitz constraint enforced, and the third is the APECS without the Lipschitz constraint. Each network has 5 layers with 9 neurons each, and the standard feed-forward networks use GeLU activation units. We refer to the standard feed-forward network as F, the APECS with the Lipschitz constraint as APECS, and the one without Lipschitz constraint as APECS-NL.

Each network was trained in three ways. First, we let the human loss weight $\gamma=0$ so that the neural networks would train to mimic the expert controller. Second, we let $\gamma=\frac{1}{2}$. Finally, we set $\gamma$ using the derived normalization value in Eq. \ref{eqn:optimal_gamma}. Each was trained with the Adam optimizer for 1k epochs. 


The input to the model was a vector of 7 inputs, where the training data where 10,000 data points sampled uniformly in the input range From the output of the expert operator, the  $\alpha$ constant was determined from
\begin{align*}
    \alpha &= -\min\left(\text{sign}(\hat{x}) \bar{x} \right),
\end{align*}
which resulted in $\alpha = 11.996$, given the expert controller and fuzzy logic system. The respective $\gamma$ was thus calculated to be $\gamma=0.998$. 

\begin{figure}[!ht]
\centering
\subfloat[Expert loss]{\includegraphics[width=0.49\linewidth]{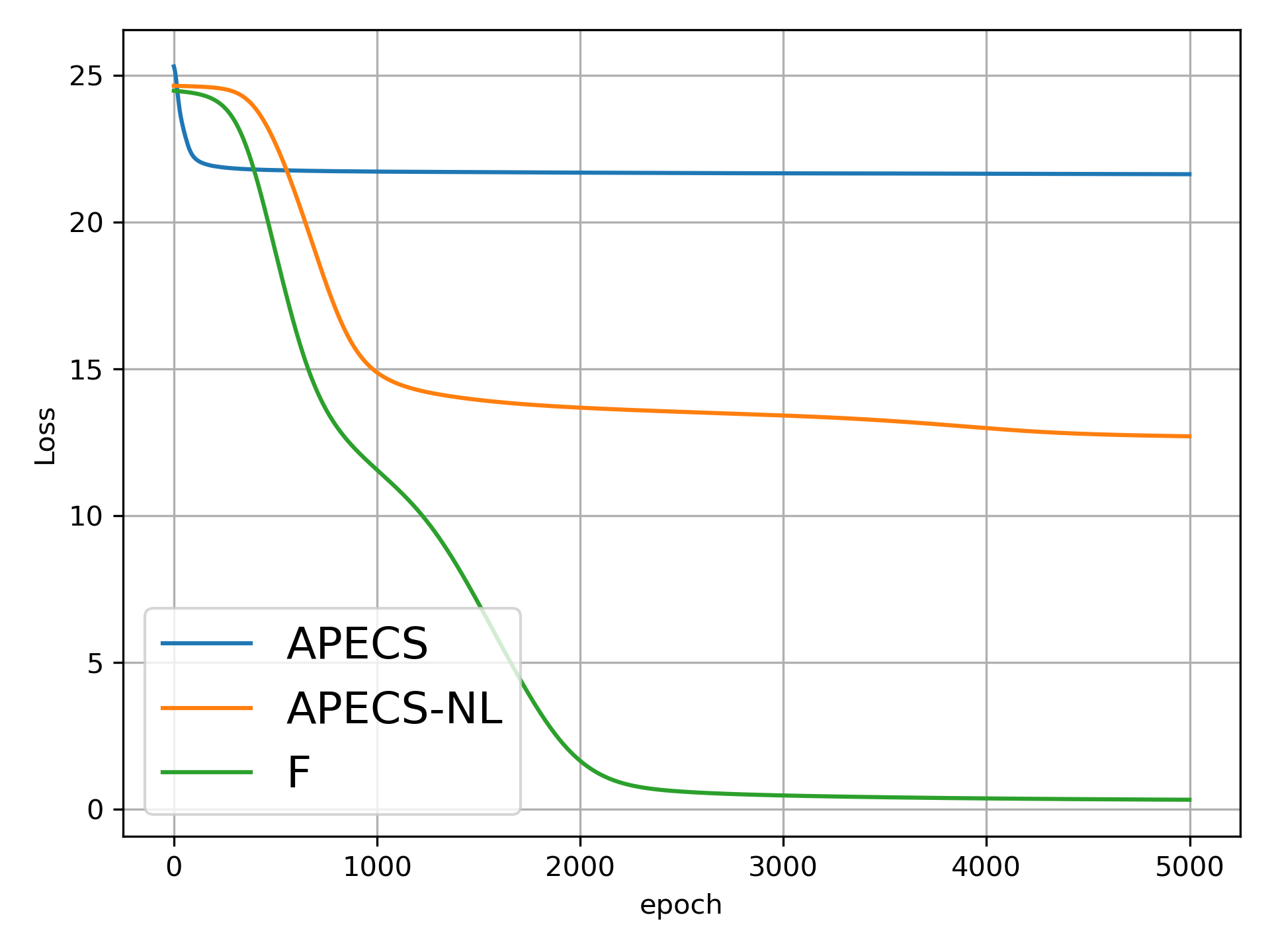}%
\label{fig:gamma_compare_0.0_l_controller}}
\hfil
\subfloat[Human loss]{\includegraphics[width=0.49\linewidth]{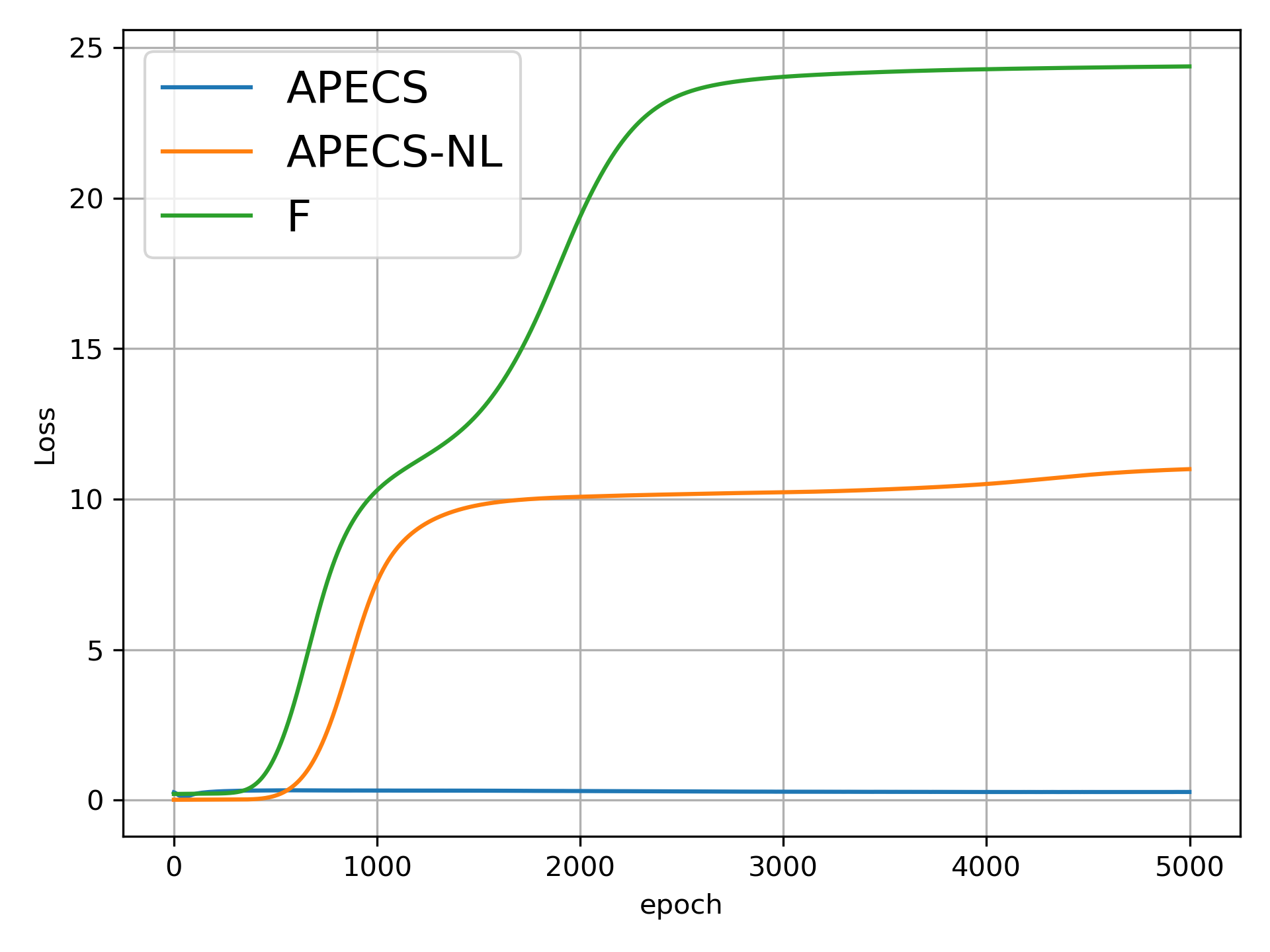}%
\label{fig:gamma_compare_0.0_l_human}}
\caption{Loss comparison for $\gamma = 0$}
\label{fig:SubGamm0}
\end{figure}

\begin{figure}[!ht]
    \centering
    \includegraphics[width=\linewidth]{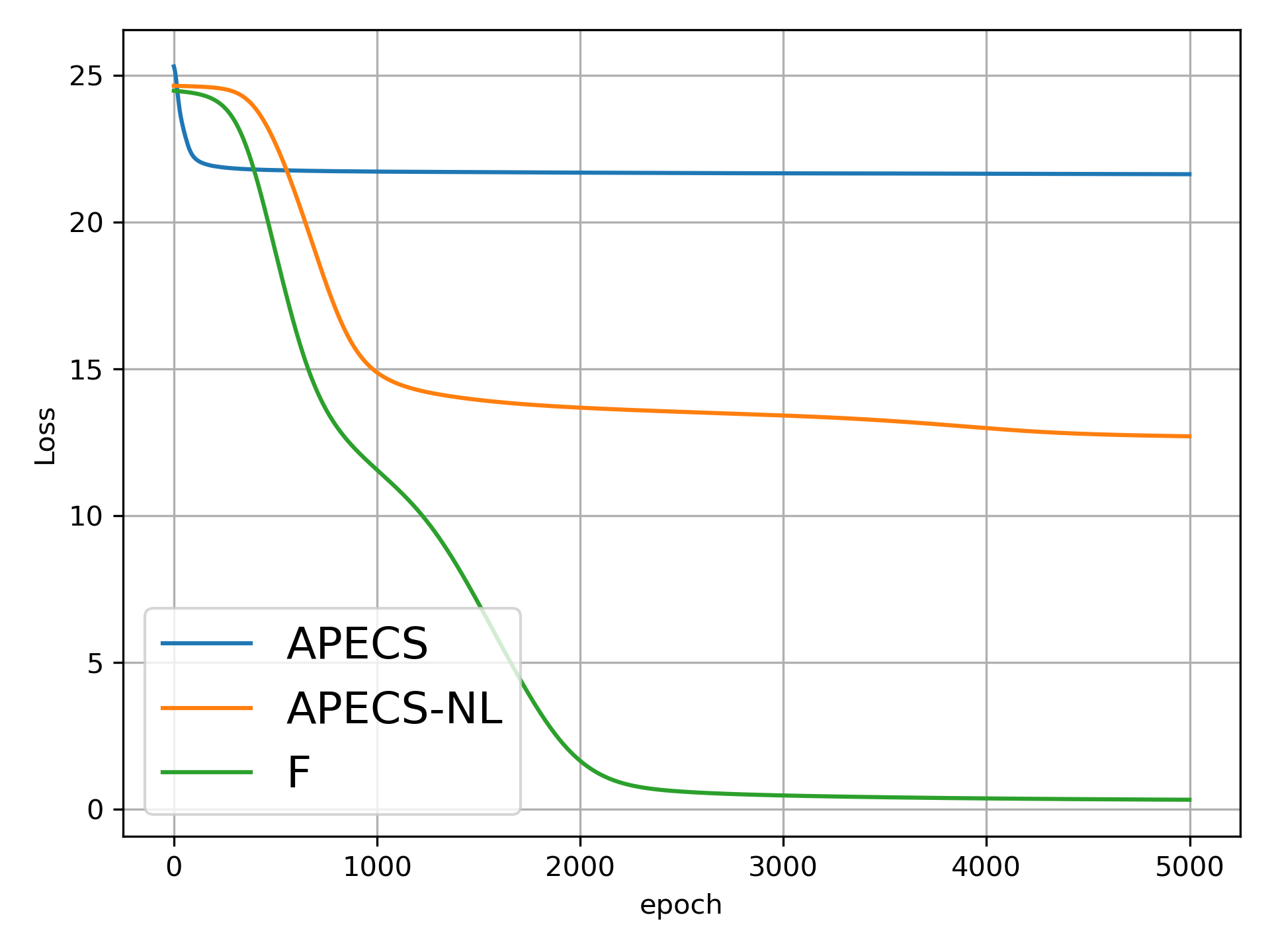}
    \caption{Total loss $\gamma = 0$}
    \label{fig:TotalGamm0}
\end{figure}

\begin{figure}[!ht]
\centering
\subfloat[Expert loss]{\includegraphics[width=0.49\linewidth]{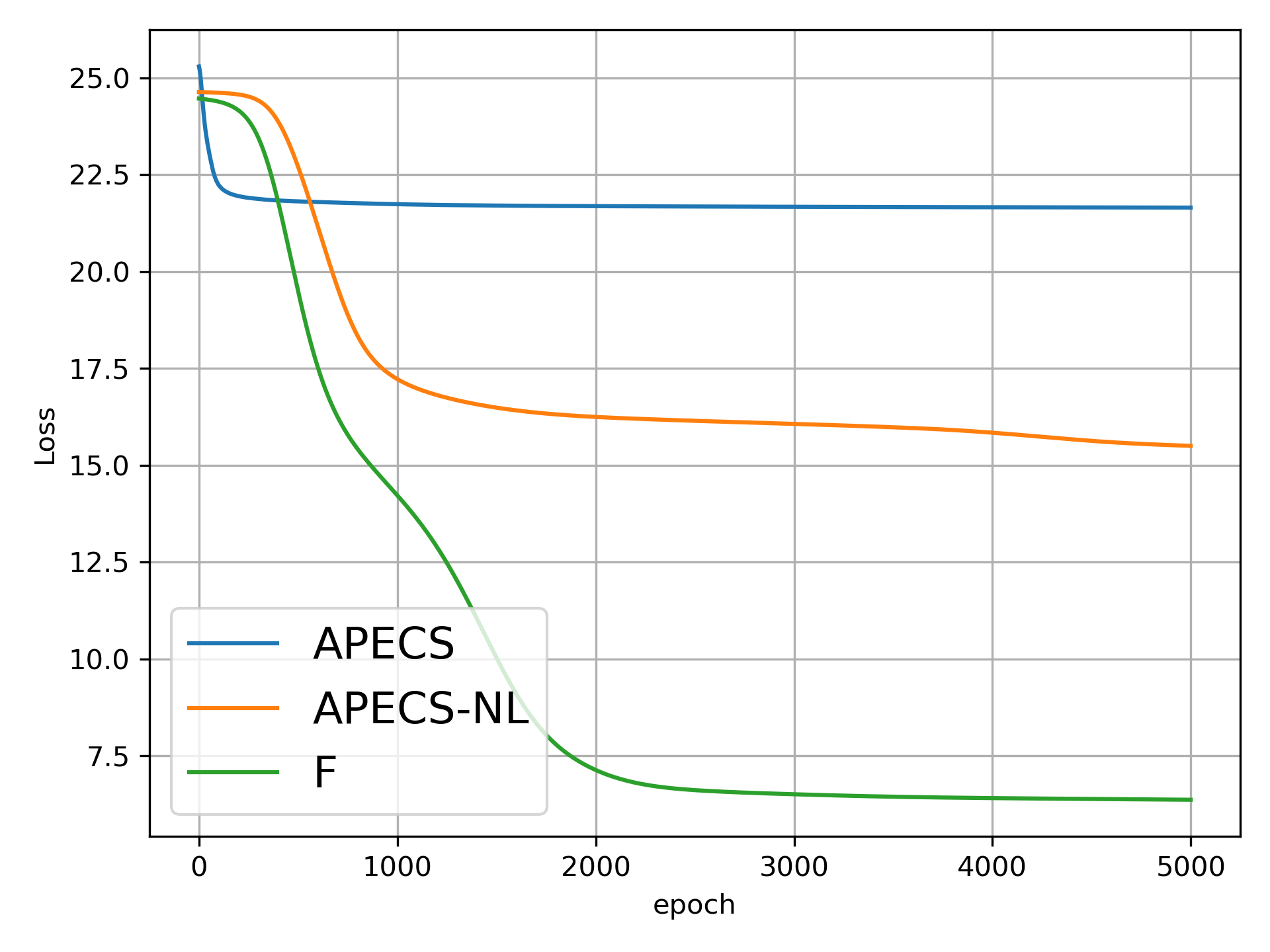}%
\label{fig:gamma_compare_0.5_l_controller}}
\hfil
\subfloat[Human loss]{\includegraphics[width=0.49\linewidth]{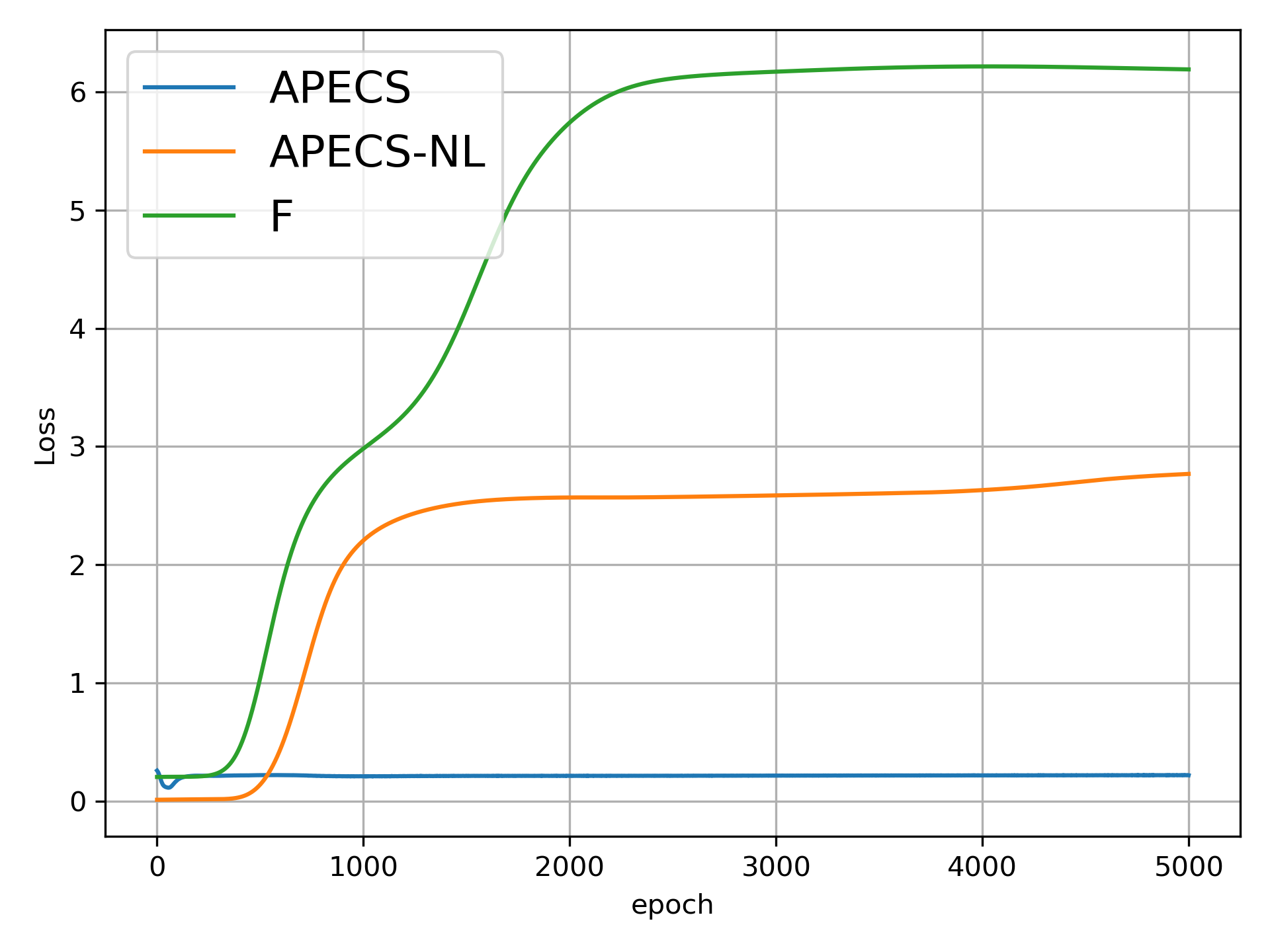}%
\label{fig:gamma_compare_0.5_l_human}}
\caption{Loss comparison for $\gamma = \frac{1}{2}$}
\label{fig:SubGamm0.5}
\end{figure}

\begin{figure}[!ht]
    \centering
    \includegraphics[width=\linewidth]{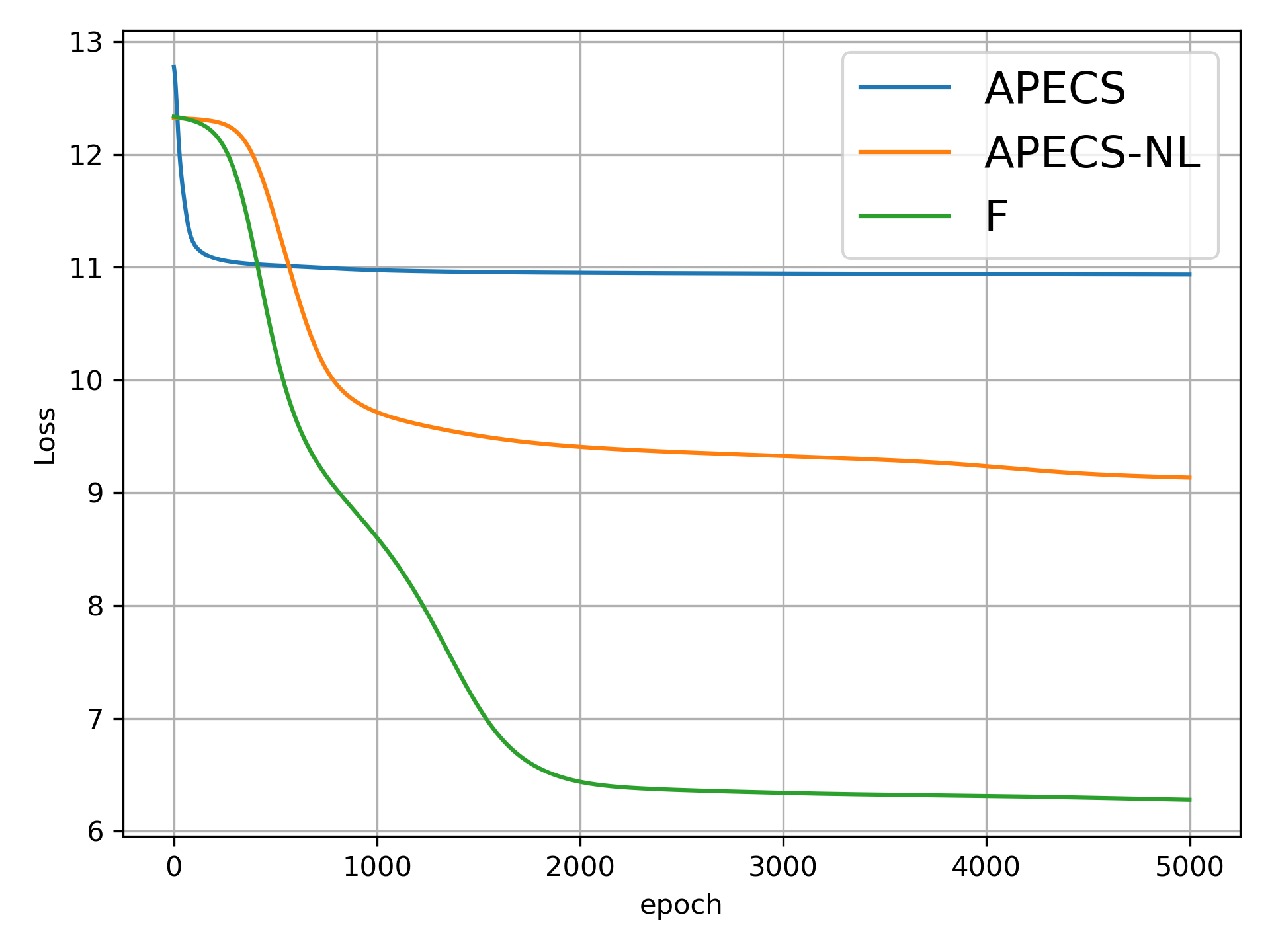}
    \caption{Total loss $\gamma = \frac{1}{2}$}
    \label{fig:TotalGamm0.5}
\end{figure}

\begin{figure}[!ht]
\centering
\subfloat[Expert loss]{\includegraphics[width=0.49\linewidth]{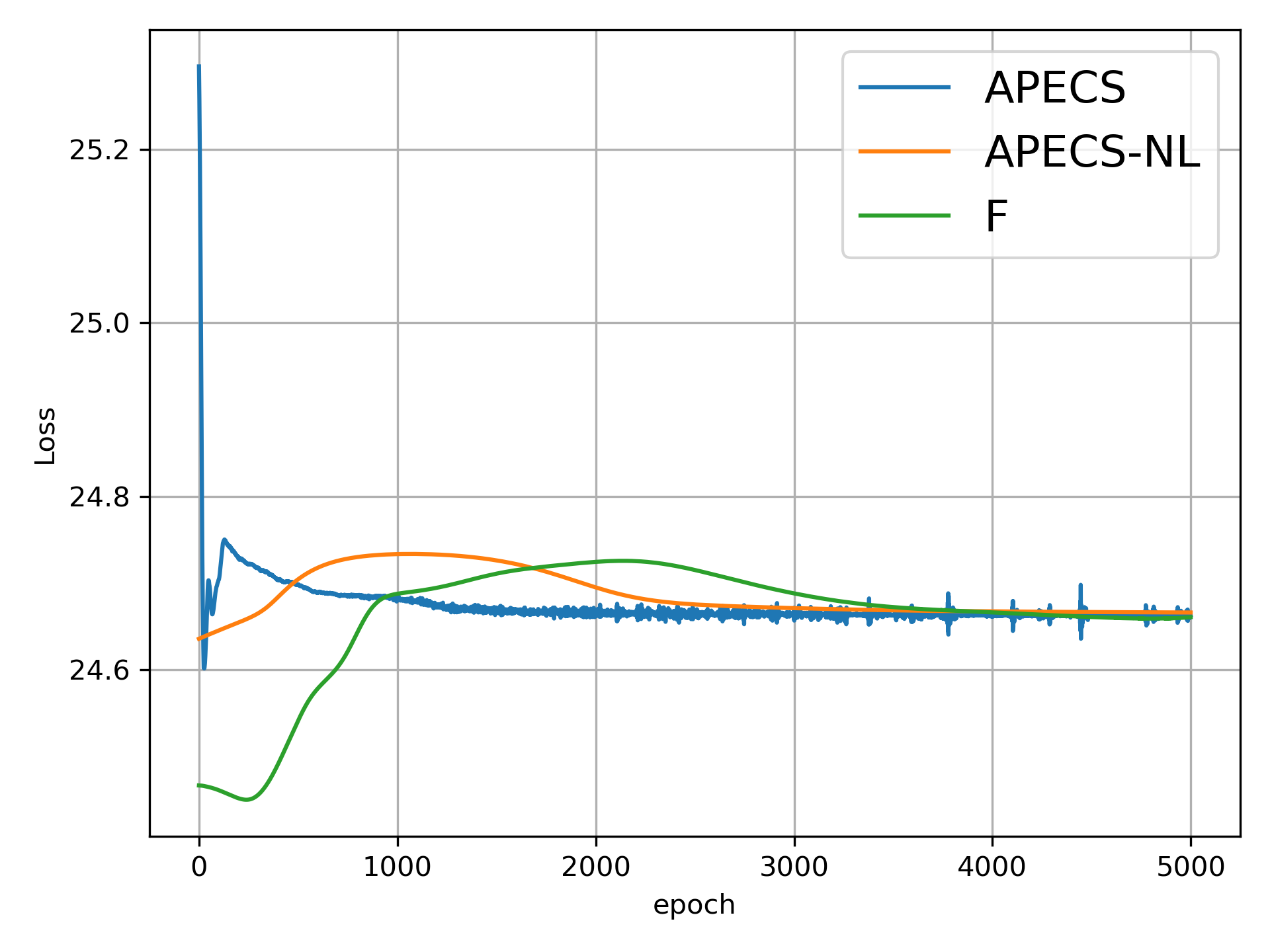}%
\label{fig:gamma_compare_0.998_l_controller}}
\hfil
\subfloat[Human loss]{\includegraphics[width=0.49\linewidth]{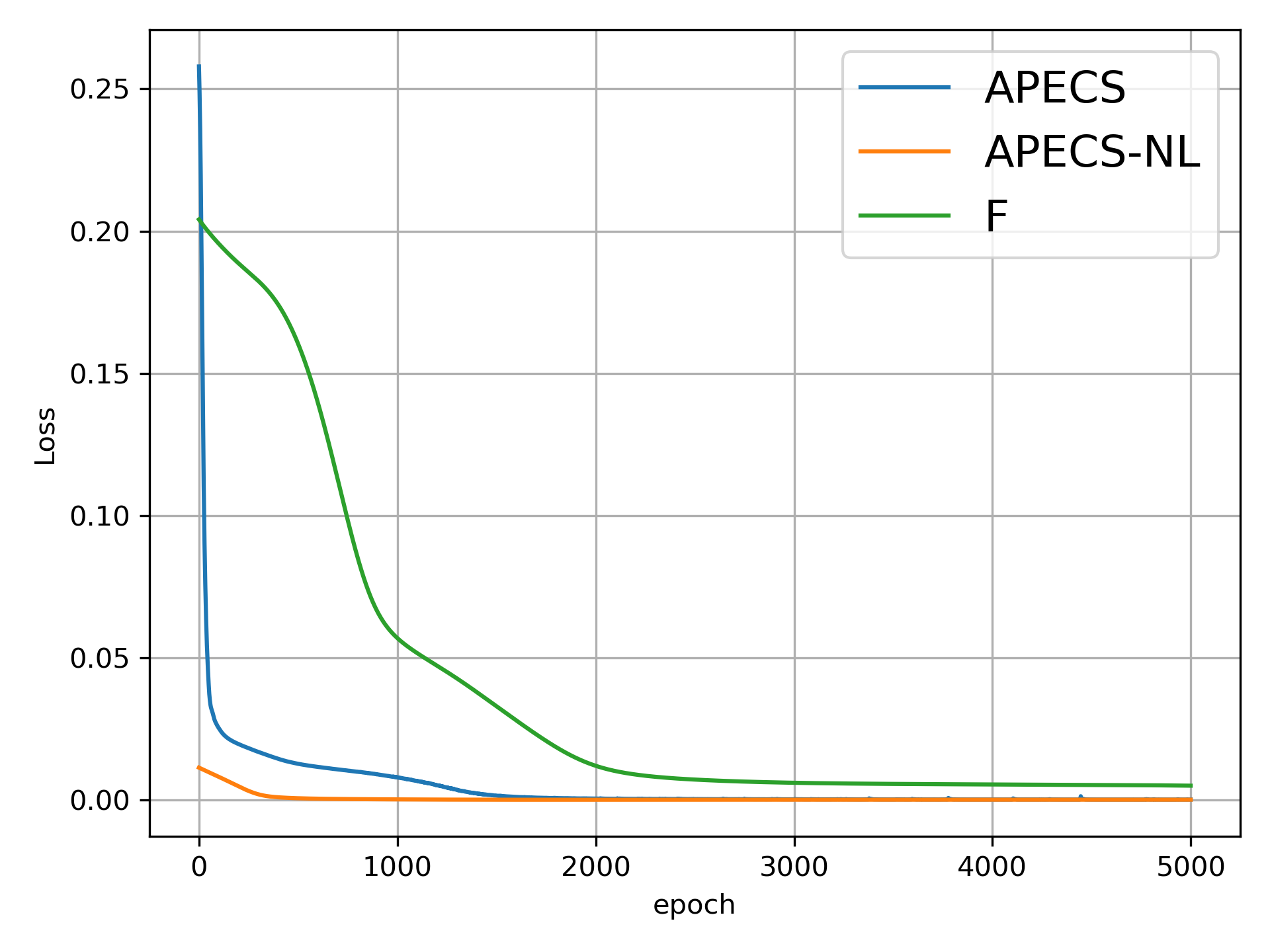}%
\label{fig:gamma_compare_0.998_l_human}}
\caption{Loss comparison for $\gamma = 0.998$}
\label{fig:SubGamm0.9}
\end{figure}

\begin{figure}[!ht]
    \centering
    \includegraphics[width=\linewidth]{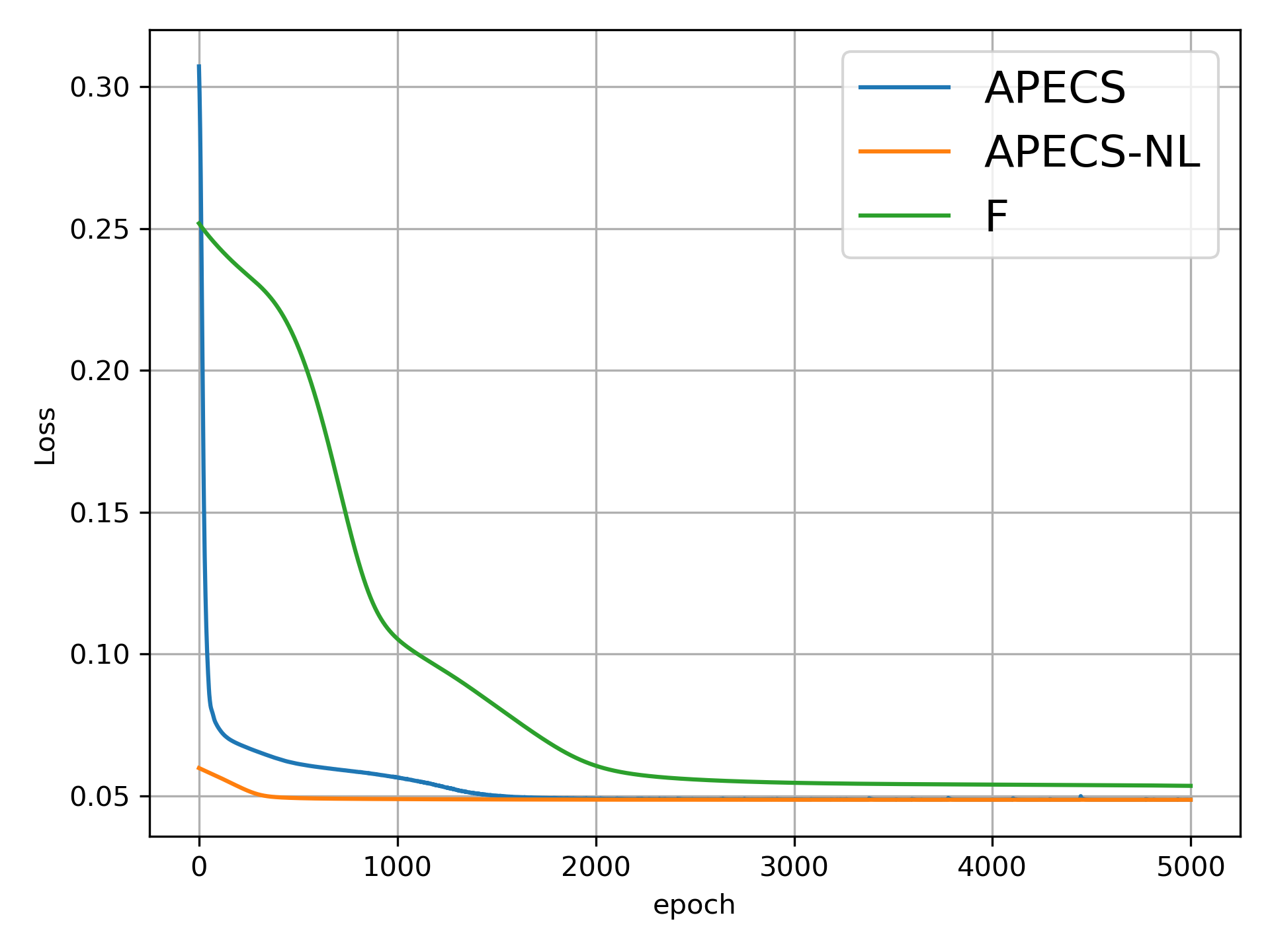}
    \caption{Total loss for $\gamma = 0.998$}
    \label{fig:TotalGamm0.9}
\end{figure}

We compare the training for different $\gamma$ values. Figures \ref{fig:SubGamm0} and \ref{fig:TotalGamm0} correspond to $\gamma = 0$. Figures \ref{fig:SubGamm0.5} and \ref{fig:TotalGamm0.5} correspond to $\gamma = \frac{1}{2}$. Figures \ref{fig:SubGamm0.9} and \ref{fig:TotalGamm0.9} correspond to the optimal $\gamma = 0.998$.

%
%
%
As demonstrated in Figures \ref{fig:SubGamm0}, \ref{fig:SubGamm0.5}, and \ref{fig:SubGamm0.9}, it is only when the derived $\gamma$ value is used that the optimizer properly minimizes both the human and expert losses; in other cases, the optimizer ignores the expert controller. 

\begin{figure}[!ht]
    \centering
    \includegraphics[width=1\linewidth]{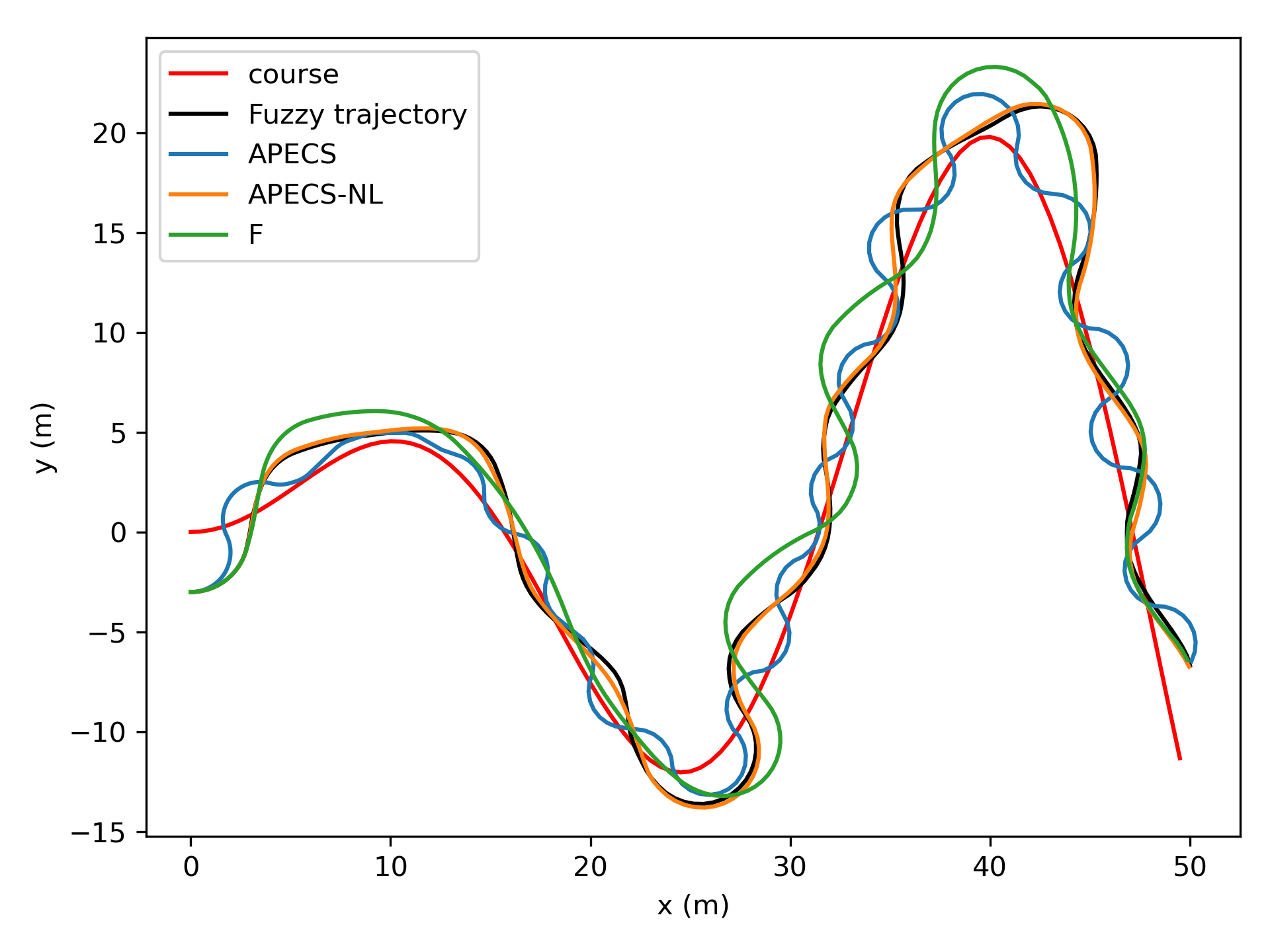}
    \caption{Comparison of controller outputs, $\gamma=0.998$}
    \label{fig:NetworkComparaison}
\end{figure}

Figure \ref{fig:NetworkComparaison} shows the trajectories taken by the controllers resulting from training with $\gamma = 0.998$. The optimized controller is the APECS controller, which follows the trajectory closer than the fuzzy logic optimizer. Both the APECS-NL and the F network structures perform worse than the APECS system, with the F network exacerbating the oscillations generated from the reference fuzzy logic output.
The root-mean-square error (RMSE) is a standard metric for measuring path-tracking ability \cite{william_norris_design_2001}. The RMSE, calculated from the perpendicular distance to the target course, for each system is displayed in Table \ref{tab:Network_RMSE_Comparisons}.
\begin{table}[!ht]
    \caption{Network RMSE Comparisons}
    \centering
    \begin{tabular}{|c|c|} \hline 
         Model& RMSE (m)\\ \hline 
        APECS & \textbf{1.473} \\ \hline
        Fuzzy & 1.542 \\ \hline
        APECS-NL & 1.592 \\ \hline
        F & 1.611 \\ \hline
    \end{tabular}
    \label{tab:Network_RMSE_Comparisons}
\end{table}
The APECS system provides a 4.5\% improvement in path tracking over the standard human controller. The characteristic oscillations, while attenuated, are maintained. It is also interesting to note that the standard neural network model F shows a performance decrease during this training, with a -4.5\% performance decrease compared to the initial human controller. Although it maintains some characteristics of the human controller output, the oscillations are very different from the actual output of the human operator, which makes them significantly smaller.

The APECS model was also trained with a varying set of Lipschitz upper bound constraints to demonstrate how the RMSE of the model results varied on different Lipschitz constraints.

\begin{figure}[ht!]
    \centering
    \includegraphics[width=1\linewidth]{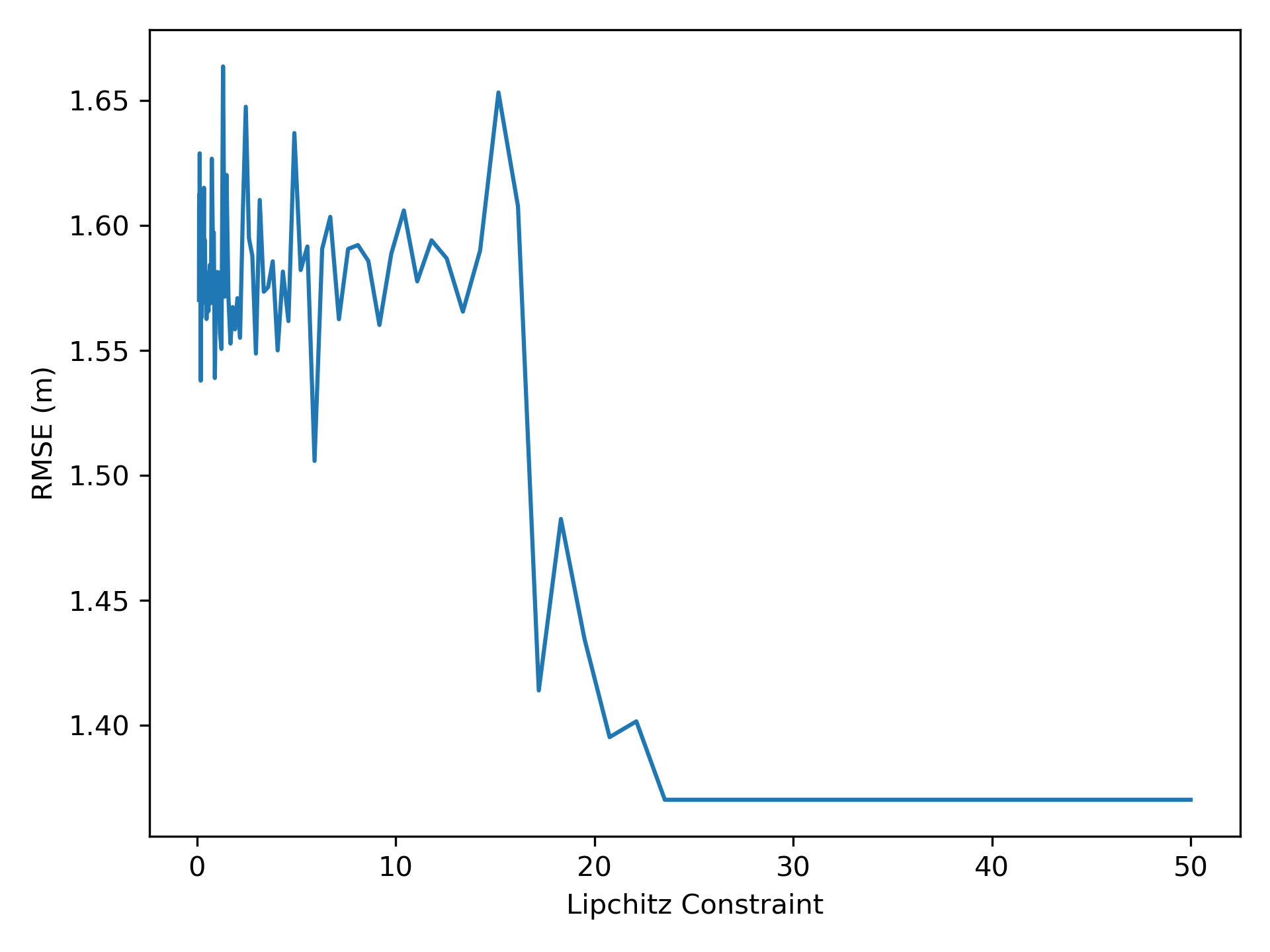}
    \caption{APECS Lipschitz constraint sweep ($\gamma = 0.998$)}
    \label{fig:LipschitzSweep}
\end{figure}

The APECS model's RMSE values were illustrated in Figure \ref{fig:LipschitzSweep}, where it demonstrated, the robustness of the network's parameterization. Over a set of varying Lipschitz values the RMSE output remained relatively constant, until it reached a higher Lipschitz value threshold of 20, where it later reached a minimum of 1.37 meters, which was a  20\% improvement from the initial Lipschitz values.

\section{Conclusion} \label{s_Conclusion}

This paper developed a novel control architecture known as the Adaptive Personalized Control System (APECS). The controller was constructed to satisfy a set of intuitively desirable properties (R1-R5). The result is a neural network based architecture which improves the control input from the human operator using an optimal controller after use of a derived training scheme. A 4.5\% performance increase compared to the initial generic fuzzy logic human operator was obtained, as well as a 9.4\% performance increase compared to using a standard neural network training to minimize the losses between mimicking the human and the expert operator.
\par
Future work will demonstrate how this architecture can be trained online to demonstrate online adaptive capabilities. In addition, during development it was noticed that the SDP 1-Lipschitz networks were sensitive to the number of layers that it had available to it, which sometimes resulted in the network instability. Similar to the works of \cite{He2015, Glorot2010}, a better initialization scheme will be explored for the 1-Lipschitz residual network to ensure that the network converges and does so efficiently.

\section{Link}

The Desmos link below demonstrates the upper bounding of the functions $p(z)x$ with a variable $g(z)$ function, with the addition of the quartic solution and the Lipschitz bounds demonstrated.
\begin{center}
\begin{small}
    \texttt{\href{https://www.desmos.com/calculator/qxr2gwxly5}{https://www.desmos.com/calculator/qxr2gwxly5.}}
\end{small}
\end{center}


\bibliographystyle{IEEEtran}
\bibliography{main}

\appendices

\section{$\gamma^*$ Constraint Satisfication} \label{sec:GammaConditionsOp}

Given the derived $\gamma^*$ in \ref{eqn:optimal_gamma}, we verify that for all $\alpha \in [0, 1]$ the conditions \ref{eqn:gammaConditions} are satisfied.

We first verify if $\gamma^*$ ever satisfies the second condition, $\gamma^* > \frac{3}{5}$:
\begin{align*}
   \frac{3 (\alpha +1)^2}{3 \alpha  (\alpha
+2)+4} > \frac{3}{5} \\
\alpha > 1 - \frac{\sqrt{2}}{2}.
\end{align*}
The second condition does indeed get satisfied as when $\alpha > 1 - \frac{\sqrt{2}}{2}$, $\gamma^* > \frac{3}{5}$ and given that is a quadratic equation with it's minimum at $\alpha = -1$, then $0 < \gamma^* < \frac{3}{5}$ with $\alpha \leq 1 - \frac{\sqrt{2}}{2}$. So the $\gamma$ requirement for the first condition is satisfied; however, it is still necessary to verify that $\alpha <\frac{2 \gamma^* }{3-3  \gamma^* }$:
\begin{align*}
    \alpha &<\frac{2 \gamma^* }{3-3 \gamma^* } \\
    & = \frac{6 (\alpha +1)^2}{(3 \alpha  (\alpha +2)+4) \left(3-\frac{9 (\alpha +1)^2}{3 \alpha  (\alpha +2)+4}\right)} \\
    & = 2(1 + \alpha)^2  \\
    0 & < 2(1 + \alpha)^2  - \alpha = 2 + \alpha (3 + 2 \alpha).
\end{align*}
Given that $\alpha \ge 0$ then $2 + \alpha (3 + 2 \alpha) > 0 $. In turn, the first condition is also always satisfied. As such, using the gamma function defined above, it is always possible to compute an initial estimate for the $L_t$ value. 

\section{Neural network initialization proof} \label{sec:y_z_equal_one}

This section goes through the derivation of the function solution such that $y(x) = x$, solving for $g(x)$.
\begin{small}
\begin{align*}
x = \frac{x \left(\sqrt{B+g(x)^2}+g(x)\right)}{2 \sqrt{\frac{1}{4} x^2 \left(\sqrt{B+g(x)^2}+g(x)\right)^2+1}} \\
    1 = \frac{\sqrt{B+g(x)^2}+g(x)}{2 \sqrt{\frac{1}{4} x^2 \left(\sqrt{B+g(x)^2}+g(x)\right)^2+1}} \\
    \sqrt{B+g(x)^2}+g(x) = 2 \sqrt{\frac{1}{4} x^2 \left(\sqrt{B+g(x)^2}+g(x)\right)^2+1} \\
    \left(\sqrt{B+g(x)^2}+g(x)\right)^2 = x^2 \left(\sqrt{B+g(x)^2}+g(x)\right)^2+4 \\
    \left(\sqrt{B+g(x)^2}+g(x)\right)^2 (1 - x^2) =  4 \\
    \left(\sqrt{B+g(x)^2}+g(x)\right)^2 =  \frac{4}{1 - x^2} \\
    \sqrt{B+g(x)^2}+g(x) =  \frac{2}{\sqrt{1 - x^2}} \\
    \sqrt{B+g(x)^2} =  \frac{2}{\sqrt{1 - x^2}} -g(x) \\
    B+g(x)^2 = \left( \frac{2}{\sqrt{1 - x^2}} -g(x) \right)^2 \\
    B+g(x)^2 = g(x)^2 - 2 \frac{2}{\sqrt{1 - x^2}} g(x) + \frac{4}{1 - x^2}  \\
    B =  - 2 \frac{2}{\sqrt{1 - x^2}} g(x) + \frac{4}{1 - x^2}  \\
     \frac{4}{\sqrt{1 - x^2}} g(x) =   \frac{4}{1 - x^2} - B  \\
     g(x) =  \frac{\sqrt{1 - x^2}}{4}  \left( \frac{4}{1 - x^2} - B  \right) \\
     g(x) =  \frac{\sqrt{1 - x^2}}{4}  \left( \frac{4}{1 - x^2} - B  \right) \\
    g(x) =  \frac{1}{\sqrt{1 - x^2}}  - \frac{B \sqrt{1 - x^2}}{4} \\
    g(x) =  \frac{4 + B \left(1 - x^2\right)}{4 \sqrt{1 - x^2}}.
\end{align*}
\end{small}

\end{document}